\documentclass[prl,aps,amssymb,showpacs,twocolumn]{revtex4}
\usepackage{amsmath}
\usepackage{amssymb}
\usepackage{amsthm}
\usepackage{amsfonts}
\usepackage{enumerate}
\usepackage{latexsym}
\usepackage[dvips]{graphicx}

\newcommand{\beq}{\begin{equation}}
\newcommand{\eneq}{\end{equation}}
\def\avg#1{\langle#1\rangle}

\def\be{\begin{equation}}       \def\ee{\end{equation}}
\def\bea{\begin{eqnarray}}      \def\eea{\end{eqnarray}}

\input{epsf}

\begin{document}

\tolerance 10000

\newcommand{\vk}{{\bf k}}


\title{ Spin Polarization and Dichroism Effects by Electric Field}
\author {  Jiangping Hu  }

\affiliation{  Department of  Physics, Purdue University,
West Lafayette, IN 47907\\
        }
\begin{abstract}

We show that electric field can induce spin polarization and
dichroism effects in angle resolved photoemission spectroscopy
(ARPES) in spin orbit coupling systems.  The physical origin behind
the effects   essentially is the same  as  the spin Hall effect
induced by the electric field. Since the ARPES experiments have both
energy and momentum resolutions,  the spin Hall effect can be
directly verified by the ARPES experiments for individual band even
if there is no net spin current.

\end{abstract}


\maketitle

The field of spintronics, which  manipulate the spin degree of
freedom in solid state devices, has been an active field of
research. One important research in this field is to  create,
control and detect spin current. Recently, a new spin current source
has been suggested\cite{Murakami2003,Sinova2004}.  It is proposed
theoretically that a spin Hall current can be generated in strongly
spin-orbit coupling systems by external electric field. The spin
Hall effect exists in a broader class of spin-orbit coupling models.
It has been evolved into a subject of intense theoretical
research\cite{Hu2003,Bernevig2004a,Culcer2004a,Rashba2003,Murakami2004a,
Murakami2004b,Murakami2004c,Rashba2004,Bernevig2004b,Burkov2004a,Chang2004,Hirsch2004,
Schliemann2003,Schliemann2004a,Sinitsyn2004a,Shen2004a,Sheng2004b,Liu2004}.

The spin Hall effect can be easily derived from the single particle
quantum mechanics. In real materials, due to disorder, theoretically
it is still controversial  whether the effect exists or not.
Recently, two experimental groups \cite{Kato2004,Wunderlich2004}
have reported that the spin Hall effect has been observed in  two
dimensional hole systems with Rashba spin-orbit coupling, which
contradicts  the theoretical analysis in the presence of
disorder\cite{Mishchenko2004,Ioue2004}. Therefore, independent
experiment is still required to verify the effect.

In this paper, we propose that   angle resolved photoemission
spectroscopy(ARPES) can be used to detect the spin Hall effect in
spin-orbit coupling systems.  We show that electric field can induce
spin polarization and dichroism effects in angle resolved
photoemission spectroscopy (ARPES) in spin orbit coupling systems
without any magnetization. The effects stem from the same physical
origins as the spin Hall effect. The ARPES is a very powerful tool
to study condensed matter materials, such as
cuprates\cite{Damascelli2003}. Compared to other experimental
techniques, there are several advantages. First, the ARPES
experiments have both energy and momentum resolutions which provide
the detailed electronic physics of  individual band. In particular,
we will show that it is not required to have spin current flowing in
samples in order to verify the spin Hall effect. Therefore, there
could be no magnetization at the edges of samples. The signal is
induced purely by electric field. Secondly, the ARPES has been used
to measure the spin-orbit coupling\cite{Mizokawa2001,Rotenberg1999}.
The experimental setup discussed here is straightforward.  Finally,
there are several independent quantities which can be measured to
test the physics of the spin Hall effect.

Before  discussing  ARPES measurements, let's consider the original
physics of the spin Hall effect. The result that we want to
emphasize is that, in general,  there is dissipationless spin Hall
current in each band in spin-orbit coupling systems even if the
total net spin Hall current is zero. Imagining two bands which is
split by the spin-orbit coupling from a spin-degenerated band, there
is no net spin current created by external electric field if both
bands are completely filled. However, in each band, there is a spin
Hall current. The net spin Hall current is zero because the
contributions from two filled bands cancel each other. This result
is obvious if one follows the argument that the spin Hall current is
not generated  by the displacement of  electron distribution
function, but by  anomalous velocity due to the Berry curvature of
Bloch states\cite{Murakami2003,Hu2003}. The consequence from this
picture is very important to experimental techniques such as the
ARPES which can access the physics of individual band.

To show the above analysis, let's follow the formulism given in
ref.\cite{Hu2003}. Let us consider a general spin-orbit coupling
model described by  Hamiltonian, $H( P, S)$. In the presence of a
constant external electric field, we choose  vector potential, $\vec
A=-\vec E t$. The total Hamiltonian becomes time dependent,
$H(t)=H(P-e\vec E t,  S)$. Let $ |G, t>$ be an instantaneous ground
state  of the time-dependent Hamiltonian,
\begin{eqnarray}
H(t) |G,t>=E_{G}(t)|G, t>.
\end{eqnarray}
By first-order time-dependent perturbation theory, we have
\begin{widetext}
\begin{eqnarray}  \label{kubo}
&&|\Psi_{G}(t)\rangle=\exp\{ -i \int_0^t dt^\prime E_G(t^\prime)\}
\Big \{ |G,t \rangle + i \sum_n \frac{|n,t\rangle \avg{n,t|
\frac{\partial}{\partial t} |G,t}} {E_n(t)-E_G (t)} (1-e^{i(E_n(t)-
E_G(t))t}) \Big \},
\end{eqnarray}
\end{widetext}
where $|n,t\rangle$ are   excited instantaneous eigenstates. Now,
let many body ground state be two bands split by spin orbital
coupling. The ground state wavefunction for each band is given by
$|G_1> = \prod_{k<k_{F1}}|k,\lambda_1>$ and $|G_2> =
\prod_{k<k_{F2}}|k,\lambda_2>$ respectively, where
$|k,\lambda_{1,2}>$ label   single particle states. Thus, in
adiabatic approximation in the presence of electric field $\vec E$,
we have  new ground state wave functions $|G_1(\vec E)>$ and
$|G_2(\vec E)>$,
\begin{widetext}
\begin{eqnarray}
& & |G_1(\vec E)> =|G_1>+ ie\vec E \cdot \sum_{k<k_{F1}} \vec B(\vec
k)
|k,\lambda_2> \prod_{ k'\neq k, k'<k_{F1}}|k',\lambda_1> \nonumber \\
& & |G_2(\vec E)> =|G_2>- ie\vec E \cdot \sum_{k<k_{F2}} \vec
B^*(\vec k) |k,\lambda_1> \prod_{ k'\neq k, k'<k_{F2}}|k',\lambda_2>
\label{wavefunction}
\end{eqnarray}
\end{widetext}
where $\vec B(\vec k)$ is given by $\vec B(\vec k)
=\frac{1}{\Delta(k)}<k,\lambda_2|\frac{\partial}{\partial_{\vec
k}}|k, \lambda_1>$, and $\Delta(k)$ is   spin-orbit splitting
energy. The adiabatic approximation is valid when $\frac{\hbar
eE}{k\Delta(k)}<< 1$. $\vec B (\vec k)$ is  precisely the Berry
curvature of the Bloch states and is non-vanishing in spin-orbit
coupling systems in general. The second term in
eq.\ref{wavefunction} is responsible for the spin Hall
effect\cite{Hu2003}. From the wavefunctions,  it is clear that the
spin Hall current is contributed by all the particles in the bands.
Even if the two bands are completely filled, the physics of the spin
Hall effect still exists in each band although the total spin
current is zero because the spin currents in the two bands run in
opposite directions with equal amplitude and cancel each
other\cite{Murakami2003,Hu2003}. Therefore, we can detect the spin
Hall effect independently if we can manage to observe the second
part of wavefunctions. Moreover, the detection can be done even in
completely occupied bands if the individual band can be access
separately. Modern photoemission experiments have achieved
remarkable energy and momentum resolution. It should be an ideal
technique to measure such effects.

In the photoemission experiment, the transition probability between
an initial state $\Psi_i(N)$ and final state $\Psi_f(N)$ is given by
\begin{eqnarray}
\Omega_{fi}=
\frac{2\pi}{\hbar}|<\Psi_f(N)|\hat{\Delta}|\Psi_i(N)>|^2
\delta(\hbar\omega_f-\hbar\omega_i-\hbar\omega)
\end{eqnarray}
where $\hat{\Delta}$, under dipole approximation, is given by
\begin{eqnarray}
\hat{\Delta}=\frac{e}{2mc}(\hat{A}\cdot\hat{P}+\hat{P}\cdot\hat{A})=\frac{e}{mc}\hat{A}\cdot\hat{P}
\end{eqnarray}
where $\hat A$ is the electromagnetic vector potential and we have
assumed that  $\bigtriangledown \cdot \hat A=0$, and  $\hat{P}$ is
the electron momentum operator.   The total intensity is obtained by
sum over all the initial and final states in the system. In
principle,  the photoemission experiments can provide information
specified by four quantities,  including   energy $\omega$, momentum
$\vec k$, spin $\sigma(\vec{m})$ which is associated to a specified
direction $\vec m$ and  the polarization of photon source, $h$. The
total intensity of photoelectrons can be viewed as a function of the
above quantities, i.e. $I^{h}(\omega,\vec k, \sigma_{\vec m})
=\sum_{fi}\Omega_{fi}$. To simplify the discussion, we would like to
take   familiar three step approximation of photoemission which is a
good approximation in general. In this approximation, the total
intensity is proportional to the product of matrix element and
single spectral function $A(\vec k, \omega)$ which provides the
information of the band structure, namely
\begin{eqnarray}
 I^{h}(\omega,\vec k,
\sigma(\vec m))\propto M^h(\vec{k},\sigma(\vec m)) A(\vec k,
\omega),
\end{eqnarray}
where    the matrix element is given by
\begin{eqnarray}
M^h(\vec{k},\sigma(\vec m))
=\sum_{fi}|<\psi_f|\hat{\Delta}|\psi_i>|^2.
\end{eqnarray}
In   spin-orbit coupling systems, the matrix element  includes  the
important information of the spin Hall effect. Let's consider the
model  discussed earlier.   In the presence of electric field,
plugging eq.\ref{wavefunction} into the above equation, we obtain
\begin{widetext}
\begin{eqnarray}
\Delta M^h(\vec{k},\sigma(\vec m),\lambda)= M^h(\vec{k},\sigma(\vec
m)) -M_0^h(\vec{k},\sigma(\vec m))= -i e\vec E\cdot \vec B(\vec
k,\lambda) \sum_{f,\lambda'\neq\lambda} Re<\psi_f|\hat{\Delta}|\vec
k_i,\lambda><\psi_f|\hat{\Delta}|\vec k_i,\lambda'>^*).
\label{result}
\end{eqnarray}
\end{widetext}
 where  $M_0^h(\vec{k},\sigma(\vec m))$ is the matrix element
without the external electric field.  $\Delta
M^h(\vec{k},\sigma(\vec m),\lambda)$ describes the response to the
external electric field and is proportional to the Berry curvature
of Bloch states in momentum space due to the spin orbit coupling.
This is the main result of the paper. $\vec B(\vec k_i)$ is purely
imaginary. In principle, we should be able to calculate the terms in
the above equation theoretically for different materials and measure
the quantities in the photoemission experiments. In the following,
we would like to simplify the results for several spin-orbit
Hamiltonians first and discuss the qualitative measurements which
can be done to test the prediction while the detailed calculation
for different materials is left to be reported elsewhere.

Before we discuss   specific models, we would like to state the
general properties from eq.\ref{result} for a  band which is split
to two bands by spin orbit coupling: (1) $\Delta
M^h(\vec{k},\sigma(\vec m),\lambda)$ is directly proportional to the
external electric field and Berry curvature. Therefore, it carries
the sign (or direction) information of the wavevector $\vec k $; (2)
for two bands which are labeled by $\lambda =\pm $, $\Delta
M^h(\vec{k},\sigma(\vec m),+)=-\Delta M^h(\vec{k},\sigma(\vec m),-)$
which reflects the same nature of the spin  Hall current, namely,
the spin Hall currents are exactly opposite in two bands;
(3)defining
 \begin{eqnarray} g(\vec k)=\sum_{f}Re
<\psi_f|\hat{\Delta}|\vec k,+><\psi_f|\hat{\Delta}|\vec k,->^*,
\end{eqnarray} which is determined by the detailed properties of
the band. Although it is not easy to calculate $g(\vec k)$, we can
make use of its symmetry properties. We will illustrate this point
later.

 Let us now consider the  Rashba spin orbit coupling
Hamiltonian, which is given by
\begin{eqnarray}
H_R = \frac{ P ^2}{2m}+\gamma (P_x S_y-P_y S_x).
\end{eqnarray}
The eigenstates are given by
\begin{eqnarray}
|\vec k, \lambda>=U_R|\lambda >,
\end{eqnarray}
with eigenvalues given by $\epsilon(\vec k,\lambda)=\frac{\hbar^2
k^2}{2m}+\gamma\lambda \hbar |\vec k|$, $\lambda=\pm$.
$U_R=e^{-i\phi S_z}, \phi=\tan^{-1}\frac{k_y}{k_x}$. We obtain
\begin{eqnarray}
\Delta M^h(\vec{k},\sigma(\vec m),+)= \frac{ e\epsilon_{ij}
E_ik_j}{\gamma k^3}\cdot
  g(\vec k) \label{result1}
\end{eqnarray}
where $\epsilon_{ij}$ is a rank-2 antisymmetric tensor. If there is
a Dresselhaus spin orbit coupling term due to the lack of inversion
symmetry in bulk, which is given by \bea H_d=\beta(P_x S_x-P_yS_y).
\eea The result for $\Delta M^h(\vec{k},\sigma(\vec m),+)$ in the
presence of both spin orbit coupling terms is given by
\begin{eqnarray}
\Delta M^h(\vec{k},\sigma(\vec m),+)= (\gamma^2-\beta^2)\frac{
e\epsilon_{ij} E_ik_j}{\epsilon_{rd}(k)^3}\cdot
  g(\vec k) \label{result0}
\end{eqnarray}
where $\epsilon_{rd}(k)=k\sqrt{(\gamma^2+\beta^2)+2\gamma\beta
sin(2\phi)}$.

For the Luttinger spin orbit coupling model\cite{Murakami2003},
which is given by
\begin{eqnarray}
H_L(t)=\frac{1}{2m}[(\gamma_1+\frac{5}{2}\gamma_2)( P)
^2+2\gamma_2(P\cdot S)^2].
\end{eqnarray}
For a given $\vec P =\hbar \vec k$, the Hamiltonian has four
eigenstates,
\begin{eqnarray}
&& H|k,\lambda>=\epsilon_{\lambda}(k) |k ,\lambda\rangle,
\nonumber \\
&& \frac{\vec k \cdot S}{|k |}|k ,\lambda>=\lambda|k ,\lambda
\rangle.
\end{eqnarray}
where $\epsilon_{\lambda}(k)=\frac{\hbar^2k^2}{2m}(\gamma_1+
(\frac{5}{2}-2\lambda^2)\gamma_2)$. For $\lambda=\pm \frac{3}{2}$
and $\lambda=\pm \frac{1}{2}$, they are referred to as the heavy
hole band and light hole band respectively. Without losing the
generality, we set the electric field along $z$ direction. For the
light hole band, we obtain
\begin{eqnarray}
\Delta M^h(\vec{k},\sigma(\vec m),\pm\frac{1}{2})= \frac{ eE
\sqrt{(k_x^2+k_y^2)}}{\gamma_2 k^4}\cdot
  g(\vec k) \label{result2}
\end{eqnarray}

\begin{figure}
\includegraphics[width=5cm, height=4cm]{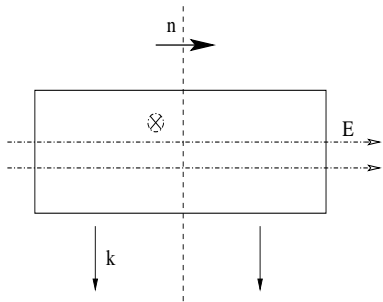}
\caption{\label{fig1} The experimental setup for detecting spin
polarization and dichroism. The electric field, $E$,  is normal to
the Mirror plane. Both measured  direction  $k$ and  the light are
in the mirror plane.}
\end{figure}

  $\Delta M^h(\vec{k},\sigma(\vec m),\lambda)$ is the change of
  matrix elements induced by external electric field. Instead of measuring the direct value of the change,
  which is rather difficult to do experimentally, we propose  several quantities which  are relatively easier to be
  measured. We consider a lattice with a mirror plane.
  Let $\vec{n}$ denote the mirror plane of the lattice and  $\hat{R}_{\vec{n}}$ be the reflection
operator associated to it. For a general purpose, we also consider
polarized light. If the direction of the light is in the mirror
plane, we have the following identity,
\begin{eqnarray}
\hat{R}_{\vec{n}}\Delta^r \hat{R}_{\vec{n}}^{-1}=\Delta^{l},
\end{eqnarray}
where $\Delta^{r,l}$ denote the dipole coupling operators for the
right and left polarized photon sources respectively. Under the
reflection, the final state $\psi_f(\vec k,E,\sigma(\vec m))$ is
changed to
$\psi_f(\hat{R}_{\vec{n}}k,E(\hat{R}_{\vec{n}}k),\hat{R}_{\vec{n}}\sigma(\vec
m))$. There are several special cases. $\hat{R}_{\vec{n}} \vec k=
\vec  k$ when $k$ is  in the mirror plane and $\hat{R}_{\vec{n}}\vec
k=-\vec k$ when $\vec k$ is perpendicular to the mirror plane.
$\hat{R}_{\vec{n}}\sigma(\vec m)=\sigma(\vec m)$ when the direction
for the spin measurement, $\vec{m}$, is perpendicular to the mirror
plane and $\hat{R}_{\vec{n}}\sigma(\vec m)=-\sigma(\vec m)$ when it
is in the mirror plane. Associated with these special cases, we can
define the corresponding quantities,  $D_i$, as follows:\\
(a).
$\vec k, \vec{m}$ is in the mirror plane:
\begin{eqnarray}
D_0(\vec k,\sigma(\vec m))=I^r(\vec k,\sigma(\vec m))-I^l(\vec
k,-\sigma(\vec m));
\end{eqnarray}
(b). $\vec k$ is in the mirror plane  and $\vec{m}$ is perpendicular
to the mirror plane:
\begin{eqnarray}
D_1(\vec k,\sigma(\vec m))=I^r(\vec k,\sigma(\vec m))-I^l(\vec
k,\sigma(\vec m));
\end{eqnarray}
(c). $\vec{m}$ is in the mirror plane and $ k$ is perpendicular to
the mirror plane:
\begin{eqnarray}
D_2(\vec k,\sigma(\vec m))=I^r(\vec k,\sigma(\vec m))-I^l(-\vec
k,-\sigma(\vec m));
\end{eqnarray}
(d). $\vec k, \vec{m}$ both are perpendicular to the mirror plane:
\begin{eqnarray}
D_3(\vec k,\sigma(\vec m))=I^r(\vec k,\sigma(\vec m))-I^l(-\vec
k,\sigma(\vec m));
\end{eqnarray}
  Without the external electric field, the above four quantities are
  zero due to the refection symmetry. Now, we consider an
  experimental setup as sketched in fig.\ref{fig1} where the applied
  electric field is normal to the mirror plane.
 All of the above
  four quantities are proportional to $\Delta M^h(\vec{k},\sigma(\vec
  m),\lambda)$ for each given band. In particular, we define the spin
  polarization when both $\vec k$ and
$\vec{m}$ are in the mirror plane
  \bea
P(\vec k) = \sum_{a=r,l}I^a(\vec k,\sigma(\vec m))-I^a(\vec
k,-\sigma(\vec m)) \eea and the dichroism when $\vec k$ is in the
mirror plane,
 \bea D(\vec k) = \sum_{\sigma}(I^r(\vec k,\sigma)-I^l(\vec k,\sigma)) \eea Both
$P$ and $D$ are proportional to $\Delta M^h(\vec{k},\sigma(\vec
  m),\lambda)$, i.e. \bea P(\vec k), D(\vec k) \propto \Delta M^h(\vec{k},\sigma(\vec
  m),\lambda)
\eea Since both above quantities are proportional to external
electric field, it is easy to verify whether the spin polarization
and dichroism can be induced by the electric field experimentally.
In a material without a mirror plane, it is still possible to detect
the effect by observing the change of $P$ and $D$ according to the
electric field although their values are not zero in general at zero
field.

In conclusion, we have shown that electric field can induce spin
polarization and dichroism effects in spin orbit coupling systems in
ARPES. They share the same physical origins as the spin Hall effect.
 The values of the effects  are exact opposite in the two bands
which are split by spin orbit coupling. Since the spin polarization
and dichroism effects are detected by completely different ARPES
experiments, our predictions can be independently checked, which is
crucial to resolve the present controversy regarding the existence
of the spin Hall effect.

The author would like to  thank  Z. X. Shen, D. L. Feng  and N. P.
Armitage for valuable discussions. This work is supported by Purdue
research funding.

\bibliography{spinarpes}
\bibliographystyle{h-physrev3}











\end{document}